\documentstyle[prb,aps,twocolumn,floats,epsf]{revtex}

\begin{document}


\title{Possible Coexistence of Rotational and Ferroelectric
Lattice Distortions in Rhombohedral PbZr$_{x}$Ti$_{1-x}$O$_3$} 

\author{Marco Fornari}
\address{Center for Computational Materials Science, Naval Research Laboratory,
Washington D.C. 20375 \\and Institute for Computational Sciences and Informatics, George
Mason University, Fairfax, Virginia 22030}
\author{David J. Singh}   
\address{Center for Computational Materials Science, Naval Research Laboratory,
Washington D.C. 20375}

\date{\today}
\maketitle

\begin{abstract}
The competitions between ferroelectric and rotational 
instabilities in rhombohedral PbZr$_x$Ti$_{1-x}$O$_3$
near $x = 0.5$ are investigated using first
principles density functional supercell calculations.
As expected, we find a strong ferroelectric instability. However,
we also find a substantial R-point rotational instability, close to
but not as deep as the ferroelectric one. This is similar to the situation
in pure PbZrO$_3$. These two instabilities are both strongly
pressure dependent, but in opposite directions
so that lattice compression of less than 1$\%$ is sufficient to change
their ordering. Because of this,
local stress fields due to B-site cation disorder may lead to
coexistence of both
types of instability are likely present in the alloy near
the morphotropic phase boundary.
\end{abstract}

 

Rhombohedral PbZr$_{x}$Ti$_{1-x}$O$_3$ (PZT) ceramic
alloys with compositions near
the morphotropic phase boundary (MPB) around $x = 0.52$ form the basis of most
piezoelectric transducer devices. \cite{Uchino}
This is due to a combination
of high response, realizable strains in the tenths of $\%$
range and favorable weak temperature dependencies. \cite{JaffeCook,LinesGlass}
Recently, there has been renewed scientific interest in these materials,
driven partly
by the discovery of nearly order of magnitude higher performance
in related relaxor single crystals, \cite{relaxor,E_DM_AK_DS}
and partly by the fundamental understanding of them,
being obtained by experimental investigation
and first principles calculations.

The microscopic physics underlying the low temperature phases of
the end-points is well known.
In PbTiO$_3$ ferroelectricity is due to condensation of a $\Gamma_{15}$
unstable phonon where
the oxygen octahedra shift against the cations.
The ground state structure has shifts along (001) with
a tetragonal lattice strain that stabilizes this direction.
A rhombohedral ferroelectric (FE)
phase with (111) shifts is not favored
and does not occur
because of the large electronic hybridization between Pb and O,
as may be seen by comparison with BaTiO$_3$, which
has a rhombohedral FE ground state.
\cite{Cohen_Nature1,wagh1,ss_c_k_1}

PbZrO$_3$ has a complex
anti-ferroelectric ground state
\cite{fuj1,fuj2,expPZ1,Singh_PZ} that
may be viewed as arising from the ideal cubic perovskite structure by
a combination of strong zone boundary instabilities. \cite{fuj-old,zonvan}
Significantly, the ferrodistortive
$\Gamma_{15}$ mode that gives the FE ground state of PbTiO$_3$
is also found strongly unstable in density functional (DF)
studies of cubic perovskite
PbZrO$_3$ even though the ground state is not FE.
Also the instabilities of cubic perovskite
PbZrO$_3$ are much stronger than
for PbTiO$_3$, with energies of 0.20 -- 0.25 eV per formula unit.
\cite{Singh_PZ,singh_PZKTKNb}
The actual structure
arises from a delicate balance between modes characterized as octahedral
rotations, octahedral distortions
and off-centering. With the addition of small amounts of
Ti, the PZT phase diagram shows a transition to FE behavior,
in other words freezing in of the $\Gamma_{15}$ instability. However,
the actual situation is no doubt more complex than this. First of
all, in the Zr rich part of the phase diagram, there is a low temperature
rhombohedral phase in addition to the high temperature phase characterized
by the pure $\Gamma_{15}$ displacement. This low temperature
phase is associated with a co-existence of the
frozen in $\Gamma_{15}$ FE instability with
rotations of the oxygen octahedra. \cite{JaffeCook,viehland,corker}
Secondly, local probes
indicate a complex local structure that differs
substantially from the average
diffraction structure in Zr rich PZT and also 
in the related Pb based relaxor single crystals.
\cite{E_DM_AK_DS,DM_AK_DA_Eg,egami2,teslic}

Complete phonon dispersions of cubic perovskite
PbTiO$_3$ and PbZrO$_3$ determined by DF calculations
were reported by Ghosez {\it et al.}. \cite{ghosez_PRB} Both materials
show both FE ($\Gamma_{15}$) and rotational (R$_{25}$ type) instabilities,
though in the
titanate the rotational (R$_{25}$) instability
is weak and occurs only in small regions
of the zone around the R and M points.
As mentioned, the R$_{25}$ unstable mode consists of
rotation of the oxygen octahedra
around the transition metal. The rapid upward dispersion away from R and M
is notable; it reflects the rigidity of the octahedra.
In cubic perovskite structure
PbZrO$_3$ the R$_{25}$ mode is more unstable and disperses
upwards from R weakly, implying more deformable octahedra,
a point that was associated with pressure dependencies.
\cite{coc-rabe} It should, however, be noted that in materials like
PbZrO$_3$ where the ideal structure is highly unstable, the size of 
the various instabilities is not simply related to the magnitudes of
the corresponding imaginary phonon frequencies in the cubic structure.
For example, the phonon frequencies indicate that the R and M point
instabilities associated with octahedral rotation in PbZrO$_3$ are
much more unstable than the FE $\Gamma_{15}$ mode, \cite{zonvan,ghosez_PRB}
but, energetically, the FE mode is more
unstable. \cite{Singh_PZ,singh_PZKTKNb}

Here, we use a simple ordered supercell with composition
$x = 0.5$ to investigate the relative strengths of the FE and rotational
instabilities in this region. We find that the FE instability is stronger,
as expected, but only marginally so, and that the balance between these
may be easily switched by modest strains at the sub $1\%$ level.
Noting the difference in ionic radii of Ti and Zr (and the cell volumes
of PbZrO$_3$ and PbTiO$_3$) and
that there is no evidence for cation ordering
in PZT near the MPB, \cite{order}
one may conclude, first of all,
that these two types of instability may coexist in alloys near the
MPB and, secondly, that it may be helpful to build rotational
degrees of freedom into effective Hamiltonians.

We focus on the rhombohedral side of the MPB neglecting
the rhombohedral strain that is known to be very small
in contrast to the tetragonal phase. \cite{RHOstrain} 
The supercell used to model the alloy is a 10 atom FCC cell with
alternating Ti and Zr layers along the [111] direction.
This is the same as one of the cells used by
S\'aghi-Szab\'o {\it et al.} \cite{ss_c_k_1}
to investigate piezoelectricity on the tetragonal side of
the MPB and Ramer {\it et al.} to 
investigate microscopic stress fields. \cite{rappe,rappe2}
The present DF calculations were done within the
Hedin-Lundqvist
local density approximation (LDA)
using the general potential linearized augmented plane-wave method
\cite{ds_book} with local orbital extensions to relax linearization
errors and treat semi-core states. \cite{singh-lo}
The Brillouin zone samplings were done using $4 \times 4\times 4$
special {\bf k} point meshes (note this is for the doubled
perovskite cell). Well converged basis sets of over 1650 functions
were used with sphere radii of 
2.0, 1.83, 2.25 and 1.47 $a_0$ for Zr, Ti, Pb and O, respectively.
Forces were calculated by the method of Yu {\it et al.}. \cite{yu}
\begin{table}[b!]
\caption[TAB. 1]{Calculated supercell
vibrational frequencies (cm$^{-1}$) of phonons
compatible with rhombohedral symmetry.}
\label{TAB1}
\begin{tabular}{|c|c|c|c|c|c|c|c|}  
  a$_L$ ($a_0$) &        &       &      &      &      &      &  \\\hline
 7.555      & 125i   & 16i   & 158  & 326  & 357  & 538  & 838  \\ 
 7.631      & 122i   & 33i   & 150  & 334  & 341  & 500  & 805  \\ 
 7.708      & 143i   & 64i   & 137  & 317  & 324  & 465  & 764
\end{tabular}
\end{table}
Calculations were done for three lattice parameters in $1\%$
increments, {\it i.e.}
7.555, 7.631 and 7.708 $a_0$.
While the ideal low temperature cubic perovskite structure cannot be accessed
experimentally, Jaffe {\it et al.} \cite{jaffe}
report an effective (by cell volume in the tetragonal phase) value of 7.73 $a_0$ at room temperature,
while Noheda {\it et al.} \cite{noheda} obtain 7.69 $a_0$ for both the
tetragonal (slightly above room temperature) and monoclinic
(20 K) phases at a slightly more Zr rich composition, $x = 0.52$.
Within the LDA we obtain a slightly smaller effective lattice parameter
for the supercell: 7.55 $a_0$, with cubic symmetry (but including the O
breathing), and 7.59 $a_0$ with a full relaxation into the ferroelectric
structure. Such $1 - 2 \%$ smaller lattice parameters relative to experiment
are typical of LDA errors in this class of materials. \cite{Singh_PZ,marzari}

In the ideal cubic structure, the supercell has one internal parameter
corresponding to breathing of the octahedron around Ti. At $a = 7.631$
$a_0$ relaxation of this gives an energy gain of 33 mRy (all energies
are for the 10 atom cell) with a
0.116 $a_0$ reduction of the Ti-O bond lengths -- very close to what
would be expected (0.11 $a_0$) based on the difference
in Ti and Zr ionic radii. This isotropic breathing
corresponds to the highest phonon branch as shown in Table \ref{TAB1}.
The frequency of 805 cm$^{-1}$ is higher than that calculated in
either pure PbZrO$_3$ or PbTiO$_3$
(725 cm$^{-1}$ and 612 cm$^{-1}$, respectively),\cite{ghosez_PRB}
as can be expected from
bond length considerations (note that the O is between a Zr and Ti
yielding a short relaxed Ti-O bond length).

We computed the remaining phonons compatible with a rhombohedral symmetry
(these are the $\Gamma_{15}$ and zone
folded R$_{15}$ modes of the simple cubic perovskite)
from atomic forces for a variety of small distortions about the
breathed FCC structure.
The dynamical matrix was determined by least
squares fit to these, and then diagonalized to obtain the phonon
eigenvectors and frequencies (given in Table \ref{TAB1}) for the
three volumes. Additionally, we calculated the energetics of the
simple perovskite R$_{25}$ rotational mode (which, however, is not
strictly speaking a true pure mode for the FCC supercell). We note that
the ferroelectric and rotational modes and the
$\Gamma_{15}$ and folded R$_{15}$ modes belong to different symmetries
and therefore do not mix at the harmonic level
({\it e.g.} the ferroelectric mode breaks inversion).

The 500 cm$^{-1}$ mode in Table \ref{TAB1} involves mainly O
displacements, like the breating mode:
small octahedra tilting mixed with stretching along
[001]. The other modes above 300 cm$^{-1}$ involve distortions
of the O octahedra, while the lowest stable mode involves mainly
transition metal off-centering in the octahedra.
In rhombohedral symmetry there is also one marginally unstable
and one unstable mode -- the
R$_{15}$ mode and the FE $\Gamma_{15}$ instability, respectively. 
In the ideal perovskite cell and the
true disordered alloy,
they belong to the same phonon branch and a rough interpolation suggests that
the FE instability extends over the entire Brillouin zone as expected
by comparison with the full dispersion curve in PbZrO$_3$
(Ref.\,\onlinecite{ghosez_PRB}).
\begin{figure}[t]
\epsfbox{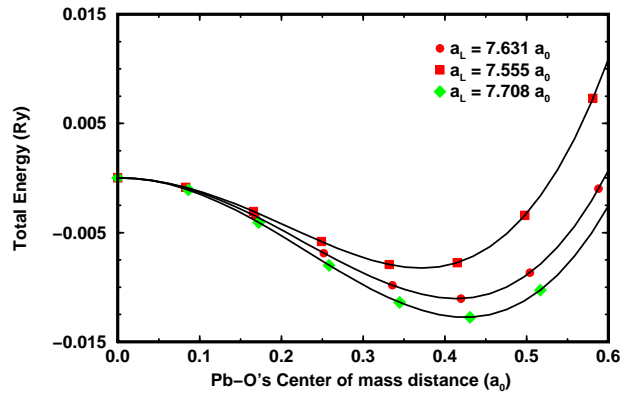}
\caption{Relaxation along the FE unstable mode (energies for
10 atom supercells) for
the reference volume (circles), 1\% compression (squares),
and 1\% expansion (diamonds). The instability decreases under pressure.}
\label{FIG1}
\end{figure}
Energy minimization along the FE
eigenmode at $a = 7.631$ $a_0$
yields an energy gain of 11 mRy (again per doubled cell).
The local minimum compatible with rhombohedral symmetry, however, occurs 
away from this line
and the relaxation provides futher 
energetic gain of 7 mRy. Its coordinates\cite{minimum} compare well   
with the displacement pattern obtained by
S\'aghi-Szab\'o {it et al.}
for the tetragonal FE state
(note that at the harmonic level the tetragonal and rhombohedral
eigenvectors are the same).\cite{ss_c_k_1}
As shown in Fig.\,\ref{FIG1}, the instability is energetically
disfavored by compression and favored by expansion. Over the range of
2\% in lattice parameter we find a variation of the energy gain along 
the FE eigenmode of about 5 mRy.
Further, the variation in depth of the local 
minima are larger: 
11, 18 and 22 mRy for $a = 7.555$ $a_0$,
$a = 7.631$ $a_0$ and $a = 7.708$ $a_0$ respectively.

The precise mechanism of the high
electromechanical response in PZT and the related relaxor crystals
is still not fully established but there are strong indications
that polarization rotation from rhombohedral to tetragonal
and the strong strain coupling for the
tetragonal direction are the main ingredients.
\cite{park,Cohen_Nature2,bellaiche} 
Recently, it has been suggested both theoretically\cite{bellaiche}
and experimentally\cite{noheda_2} that the monoclinic
phase found at the MPB\cite{noheda} can bridge the tetragonal and
the rhombohedral phase favoring such polarization rotation.
Bellaiche {\it et al.} \cite{bellaiche}
used an effective Hamiltonian approach adapted to the alloy
with parameters determined from first principles calculations.
The effective Hamiltonian used allowed strains, alloy disorder
and off-centering displacements to interact, but octahedral rotations
were not included.
Nonetheless, excellent agreement with experimental data was
obtained for the 
temperature and composition dependence of the piezoelectric
and structural properties in PZT, except
that the temperature scale had to be uniformly
adjusted downwards by approximately one third.
The microscopic mechanism that makes
polarization rotation occur at modest field strengths is still somewhat
uncertain though the importance of Pb chemistry has been widely
recognized \cite{ss_c_k_1,Singh_PZ,ghosez_PRB,bellaiche99,burton1}
In particular, it is not fully understood how the instabilities present in
PbTiO$_3$ and PbZrO$_3$ are modified in the alloy and the extent to which
they compete or perhaps coexist in real samples. 

Certainly, the simple supercell discussed above is a rather severe
approximation to the disordered alloy. Even though Zr and Ti have
the same valence, their ionic sizes and electronic properties differ
significantly. At the very least,
local stresses related to Zr-Ti disorder should be expected in PZT alloys.
Zr rich local regions will be under compressive
stress so the volume available for
the Pb ions is reduced thus lowering the local FE tendency.
Ionic considerations, supported by experimental
evidence,\cite{samara1} predict that octahedral rotations
should respond in the opposite way to compression especially
if the octahedra are stiff.
\begin{figure}[t]
\epsfbox{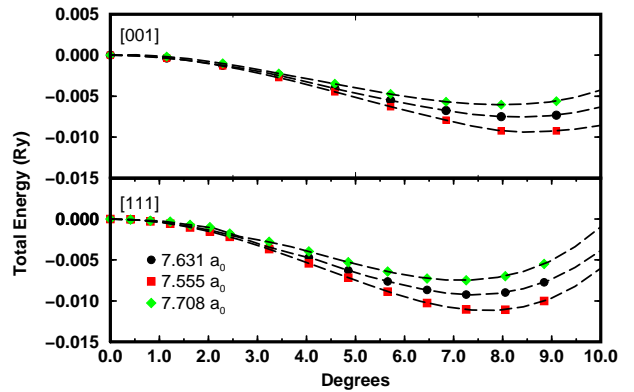}
\caption{Variation of the energy (of 10 atom supercells) with
``rotation''
(a) around [111] and (b) around [001], for
the reference lattice parameter (circles), under compression
(squares) and under tensile stress (diamonds), as in Fig.\,\ref{FIG1}.} 
\label{FIG2}
\vspace{-0.3cm}
\end{figure}
Consider such rotations around [001]
(C$_{4h}$ symmetry) and [111] (C$_{3i}$ symmetry). As mentioned, these are
not true modes for the supercell, but they are zone boundary rotational
modes that play a key role in at least PbZrO$_3$.
Both these rotations are unstable
(Fig. 2). At $a = 7.631$ $a_0$,
the energy gain for the [111] rotation is 9 mRy (7 mRy around [001]).
It is remarkable that the sizes of the FE and rotational
instabilities are so similar
for this $x=0.5$ supercell, considering that they are also
quite close, again with the FE instability lower, for PbZrO$_3$
though with a much higher energy scale.
Presumably, they track each other across the
rhombohedral side of the phase diagram. This can be qualitatively
understood in terms of ionic size effects. Unlike BaTiO$_3$ or
KNbO$_3$, \cite{ck,sb}
the ferroelectric mode in these Pb based materials
is best described as a Pb off-centering with respect to the
surrounding O with a smaller transition metal displacement in the
same direction. This is clearly seen, {\it e.g.} in the $\Gamma_{15}$
phonon eigenvector. Thus the FE instability is controlled by the
volume available to the Pb ion; the rotational mode also
involves changes in Pb-O bond length, again controlled by the
same distances.

However, a $1\%$ compression (to $a = 7.555$ $a_0$) increases the rotational
instabilities to 11 mRy about [111] and
9 mRy around [001].
Qualitatively, this is related to the stiffness of the O octahedra,
especially around Ti;
under pressure the octahedra size and geometry can only be retained
by rotation, which is made at the expense of the relatively soft
Pb-O interaction. This driving force is clearly absent for the FE
instability.
Experimental evidence of opposite stress dependencies of
rotational (R$_{25}$) and FE ($\Gamma_{15}$)
instabilities was discussed for PbZrO$_3$ early on. \cite{samara1}
Stress-field response
calculations \cite{rappe} for PbZr$_{0.5}$Ti$_{0.5}$ supercells suggest
that competition between different distortions
may be important when uniaxial stress is not
parallel to the FE distortion.
Considering the different experimental cell volumes
of the end-points (effective
$a$(PbZrO$_3$) = 7.7883 $a_0$ and $a$(PbTiO$_3$) = 7.5028
$a_0$)\cite{LB,pt_latpar}, local stresses sufficient to tip the order of the
FE and rotational instabilities seem quite reasonable in the disordered alloy.
As noted, the FE instability occurs at the $\Gamma$ and R points and probably
over most or all of the zone. Thus it can occur on a very local scale
in real space. The
rotational instability is no doubt more localized in reciprocal space, 
so several octahedra must rotate in concert within a region, but this
region may be realizably small, as the ZrO$_6$ octahedra may be soft as in
PbZrO$_3$ providing weak connections in the interlinked network.
In this scenario,
coexistence of rotational and FE distortions in the alloy is expected.
The resulting disorder might increase response away from the MPB, though
at the expense of maximum attainable response at the best composition --
this would be favorable for obtaining a desirable weak temperature dependence
of the response.
In any case, it is clear that rotational degrees of freedom
are in the low energy space along with the FE mode,
and presumably should be considered in the construction of effective
Hamiltonians, either via explicit additional coordinates or renormalization
of the existing FE and strain related coordinates. This could possibly
improve the temperature scale.

We thank R.E. Cohen, L. Bellaiche,
T. Egami and B. Burton for helpful discussions. The code
FINDSYM by H. T. Stokes and L. Boyer
was used for some symmetry analysis.
This work is supported by ONR
and the DoD ASC computer center.

\bibliographystyle{/usr/local/Revtex/prsty}

\end{document}